%% file: Lahav_Lambda_Paddy60_26March2017.tex
\documentclass[graybox]{svmult}

\usepackage{mathptmx}       % selects Times Roman as basic font
\usepackage{helvet}         % selects Helvetica as sans-serif font
\usepackage{courier}        % selects Courier as typewriter font
\usepackage{type1cm}        % activate if the above 3 fonts are
\usepackage{makeidx}         % allows index generation
\usepackage{graphicx}        % standard LaTeX graphics tool
\usepackage{multicol}        % used for the two-column index
\usepackage[bottom]{footmisc}% places footnotes at page bottom

\makeindex             % used for the subject index
\begin{document}

\title*{100 years of the Cosmological Constant: what's next?}
%: what's next?}
%past, present and future}
% Use \titlerunning{Short Title} for an abbreviated version of
% your contribution title if the original one is too long
\author{Ofer Lahav }
% Use \authorrunning{Short Title} for an abbreviated version of
% your contribution title if the original one is too long
\institute{University College London, \email{o.lahav@ucl.ac.uk}}
%\and Name of Second Author \at Name, Address of Institute \email{name@email.address}}
%
% Use the package "url.sty" to avoid
% problems with special characters
% used in your e-mail or web address
%
\maketitle

\abstract{The Cosmological Constant $\Lambda$, in different incarnations, has been with us for 100 years.
Many surveys of dark energy are underway, indicating so far that  the data are consistent with a dark energy equation of state of $w=-1$,
 i.e. a  $\Lambda$ term in Einstein's equation, although time variation of $w$ is not yet ruled out. 
The ball is now back  in the theoreticians' court, to explain the physical meaning of $\Lambda$.
We discuss sociological aspects of this field, in particular to what extent the agreement on the cold dark matter + $\Lambda$ 
concordance model 
is a result of the globalization of research and over-communication.
}

\section{Introduction}

The year 2017 marks not only that Paddy is 60 years old, but also 100 years of the Cosmological Constant $\Lambda$. 
One of the greatest mysteries in the whole of science is the prospect that 70\% of the universe is made from a mysterious substance known as `dark energy', which causes an acceleration of the cosmic expansion. A further 25\% of the universe is made from invisible `cold dark matter' that can only be detected through its gravitational effects, with the ordinary atomic matter making up the remaining 5\%
(see the Planck Collaboration 2015 study and references therein).
This ``$\Lambda$ + cold dark matter"($\Lambda$CDM) paradigm and its extensions pose fundamental questions about the origins of the universe. If dark matter and dark energy truly exist, we must understand their nature. Alternatively, General Relativity and related assumptions may need radical modifications. These topics have been flagged as key problems by researchers and by advisory panels around the world,  and significant funding has been allocated towards large surveys of dark energy. Commonly, dark energy is quantified by an equation of state parameter, $w$ , which is the ratio of pressure to density. The case $w = - 1$ corresponds to Einstein's Cosmological Constant in General Relativity, but in principle $w$ may vary with cosmic epoch, e.g. in the case of scalar fields. Essentially, $w$ affects both the geometry of the universe and the growth rate of structures. These effects can be observed via a range of cosmological probes, including the Cosmic Microwave Background (CMB), galaxy clustering, clusters of galaxies, and weak gravitational lensing, in addition to Supernovae Ia. The Hubble diagram of Type Ia Supernova (Perlmutter et al. 1999; Riess et al. 1998), for which the 2011 Nobel Prize in Physics was awarded, revealed that our universe is not only expanding but is also accelerating in its expansion. The main problem is that we still have no clue as to what is causing the acceleration, and what dark matter and dark energy are.
 Many cosmologists have puzzled over the meaning of $\Lambda$  during  the past 100 years, and it is not surprising that Paddy, 
 with his deep insight into the foundations of physics, has written many inspiring books and papers on this and related topics  (e.g. Padmanabhan 2016).

\section{Background }
\label{sec:2}

It is well known that  100 years ago Einstein added  the  Cosmological Constant $\Lambda$ to his equations in order to have a static universe (Einstein 1917)\footnote{See a historical review of this paper in O'Raifeartaigh et al. (2017)}.
His full equation is then: 
\begin{equation}
R_{\mu \nu} - {1\over 2}  R g_{\mu \nu} + \Lambda g_{\mu \nu}  = {{8 \pi G} \over {c^4}} T_{\mu \nu} \;.
\end{equation}

The big question is if $\Lambda$ should be on the left hand side, as part of the curvature, or on the right hand side, as part of the stress-energy tensor $T_{\mu \nu}$, 
for example associated with the vacuum energy $\Lambda = 8 \pi  G \rho_{vac}/c^2$. In fact a prediction for the amount of vacuum energy is expected to be $10^{120}$ times the observed value; that is a challenging problem by itself (e.g.  Weinberg 1989).

In the weak-field limit the equation of motion is:
\begin{equation}
{ d^2 r \over dt^2 } = - {{GM} \over {r^2} } + {c^2 \over 3} \Lambda r \;.
\end{equation}
A linear force was  actually already discussed by Newton in Principia in addition to the more famous inverse square law\footnote{See e.g. Calder \& Lahav (2008, 2010) for review.}.
A somewhat intuitive way to think about dark energy is as a repulsive linear force,   opposing the inverse squared gravitational force.
It is interesting that such a force can be noticeable on the Mpc scale. For example the mass of the Local Group would be estimated to be 13\% higher in the presence of a Cosmological Constant 
\footnote{See e.g. Binney \& Tremaine (2008); Partridge, Lahav \& Hoffman (2013), McLeod et al. (2016).}.
 
Should a discrepancy between data and the existing cosmological theory be resolved by adding new entities such as dark matter and dark energy, or by modifying the underlying theory? 
This reminds us of two cases in our own Solar System: the perturbed orbit of Uranus was explained by adding a new planet, Neptune, within  the existing Newtonian model.
On the other  hand, understanding the perihelion of Mercury required an entirely new theory, General Relativity \footnote{See e.g. a discussion in Lahav \& Massimi (2014) and references therein.}. 
 
There is still the possibility of another paradigm shift in our understanding of the cosmos, including the following options:

\begin{itemize}

\item{Violation of the Copernican Principle: for example, if we happen to be living in the middle of a large void;}

\item{Dark Energy being something different to vacuum energy: although vacuum energy is mathematically equivalent to $\Lambda$, the value predicted by fundamental theory is as
much as $10^{120}$ times larger than observations permit;}

\item{Modifications to gravity: it may be that General Relativity requires revision to a more
complete theory of gravity;}

\item{Multiverse: if  $\Lambda$  is large and positive, it would have prevented gravity from forming large galaxies, and life would never have emerged. Using this anthropic reasoning to explain the Cosmological Constant problems suggests a large number of universes (`multiverse') in which $\Lambda$ and other cosmological parameters take on all possible values. We happen to live in one of the universes, that is fortunately `habitable'.}

\end{itemize}

\section{The Dark Energy Survey}
\label{subsec:2}

Many ongoing and planned imaging and spectroscopic surveys  aim at measuring dark energy and other cosmological parameters. As an example we  focus here on the Dark Energy Survey (DES)\footnote{\url{http://www.darkenergysurvey.org/}}.
I have chosen  DES as it  has already accumulated data, and I happen to have been 
involved in the project since its early days back in 2004, in particular as co-chair of its Science Committee (until 2016).

DES is an imaging survey of 5000 square degrees of the Southern sky, utilising a 570 mega-pixel camera on the 4m Blanco telescope in Chile. Photometric redshifts are obtained from the multi-band photometry to produce a three dimensional map of 300 million galaxies. The main goal of DES is to determine the dark energy equation of state $w$ and other key cosmological parameters to high precision. DES will measure $w$ using four complementary techniques in a single survey: counts of galaxy clusters, weak gravitational lensing, galaxy distributions and thousands of type Ia supernovae in a `time domain' survey over 27 sq. deg. DES is an international collaboration, with more than 500 scientists from the US, the UK, Spain, Brazil, Germany, Switzerland and Australia involved. The DES science is coordinated by a Science Committee composed of eleven Science Working Groups (SWGs).
Core dark energy SWGs include large scale structure, clusters, weak lensing and supernovae Ia.
Additional SWGs focusing on the primary science are photometric \& spectroscopic redshifts,  simulations, and  theory \&  combined probes. 
The Non-cosmology SWGs focus on Milky Way science, galaxy evolution \& quasars,  strong lensing, and transients \& moving objects. 

The survey had its first light in September 2012 and started observations in September 2013. Observations are running for 525 nights spread over five years. 
The performances of several photo-\emph{z} methods applied to Science Verification data were evaluated
and the best methods yielded scatter $\sigma_{68}  = 0.08$  (defined as the 68\%  width about the median of  $\Delta z  = z_{\rm spec}  - z_{\rm phot}$). 
Regarding the image quality, the achieved median seeing FWHM is about 0.9'' in filters {\it  riz}, as  expected when designing the survey for weak lensing analyses. 
 DES has already `seen' dark matter via weak gravitational lensing (DES collaboration 2015), and the analysis towards measuring and characterising dark energy is underway.  The camera (DECam) can also capture many other celestial objects. This has resulted in both expected and unexpected discoveries (DES Collaboration 2016) including solar system objects, 17 new Milky Way companions, galaxy evolution, galaxy clusters, high-redshift objects and gravitational wave follow ups. 
 
 We highlight here two contributions of DES to  new research frontiers. Firstly, it has recently been suggested that there might be a ninth planet in the solar system. One of the six minor planets to predict `Planet 9' was discovered by DES. The DECam camera is well placed to monitor other minor planets that would help in constraining Planet 9, and of course to search for Planet 9 itself.
Secondly, the LIGO collaboration (2016) reported the first detection of gravitational waves, resulting from the merger of two black holes. This remarkable measurement confirms another of Einstein's prediction of 100 years ago. DES provided optical follow up to this event. There were no optical detections, which is not surprising, as in the conventional model a binary black hole merger is not expected to have any optical counterparts, and the DES observations covered only part of the sky where the event was likely to happen. However, DES will be vital for future LIGO follow ups. 

DES is also providing valuable experience and training of early career scientists for  on-going and future large surveys, including
the  Hyper Suprime Cam (HSC)\footnote{\url {http://www.naoj.org/Projects/HSC/}}, the Kilo-Degree Survey (KiDS)\footnote{\url{http://kids.strw.leidenuniv.nl/}},  the Large Synoptic Survey Telescope (LSST)\footnote{\url{ http://www.lsst.org/}}, \emph{Euclid}\footnote{\url{http://www.euclid-ec.org/}}, the Wide-Field Infrared Survey Telescope (WFIRST), the Subaru Prime Focus Spectrogrph (PFS)\footnote{\url{http://pfs.ipmu.jp/factsheet/}}, the Dark Energy Spectroscopic Instrument (DESI)\footnote{\url{http://desi.lbl.gov/}} and 4MOST\footnote{\url{http://www.4most.eu/}}.

\section {The globalization of research: pros and cons}

It may well be that the $\Lambda$CDM model is indeed the best description of our universe, with dark matter and dark energy ingredients that eventually will be detected independently. But there is also a chance that this is the `modern Ether' and future generations will adopt an entirely different description of the universe. It is also possible that the community has converged on a single preferred model due to `over communication'\footnote{ The discussion below is based on Lahav (2002)}.
 The society at large is going through a globalization process. There is a diversity of definitions for globalization, some in positive context, others with negative connotations. The sociologist Anthony Giddens defines globalization as ``decoupling of space and time - emphasizing that with instantaneous communications, knowledge and culture can be shared around the world simultaneously."
Another definition given in the same website sees globalization as being ``an undeniably capitalist process. It has taken off as a concept in the wake of the collapse of the Soviet Union and of socialism as a viable alternate form of economic organization."
A further discussion on globalization can be found in Thomas Friedman's book (2005) {\it The World is Flat}  (an interesting title in the context of cosmology!). He questions whether ``the world has got too small and too flat for us to adjust". 

Research in academia is of course a human activity that is affected, like any other sector, by social and technological changes and trends. The advantages of globalization to academic research are numerous: open access to data sources for all (e.g. via the World Wide Web), rapid exchange of ideas, and international research teams. These aspects make science more democratic and they enable faster achievements. The numerous conferences, electronic archives  and teleconferences generate a global village of thinkers. While
this could lead to a faster convergence in answering fundamental questions, there is also the risk of preventing independent and original ideas from developing, as most researchers might be too influenced by the consensus view.

Let us consider the above mentioned `concordance' model of cosmology. The two main ingredients, dark matter and dark energy, are still poorly understood. We do not know if they are `real' or they are the modern `epicycles'  which just help to fit the data better, until a new theory greatly simplifies our understanding of the observations. 
A disturbing question is whether the popular cosmological `concordance
model' is a result of globalization? It is interesting to contrast the present day research in cosmology with the research in the 1970s and 1980s. This was the period of the `cold war' between the former Soviet Union and the West. During the 1970s the Russian school of cosmology, led by Yakov Zeldovich, advocated massive neutrinos, `hot dark matter', as the prime candidate for dark matter. As neutrinos were relativistic when they decoupled, they moved very fast and wiped out structure on small scales. This led to the 'top-down' scenario of structure formation. In this picture `Zeldovich pancakes' of the size of superclusters formed first, and then they fragmented into clusters and galaxies. This was in conflict with observations, and cosmologists concluded that neutrinos are not massive enough to make up all of the dark matter. The downfall of the top-down `hot dark matter' scenario of structure formation, and the lack of evidence for neutrino masses from terrestrial experiments made this model unpopular. The Western school of cosmology, led by Jim Peebles and others, advocated a `bottom up' scenario,  a framework that later became known as the popular `cold dark matter'. However the detection of neutrino oscillations showed that neutrinos indeed have a mass, i.e. hot dark matter does exist, even if in small quantities.
Current upper limits from a combination of cosmic probes  are about
0.2 eV, while the lower limit from neutrino oscillations is 0.06 eV.  Therefore both forms, cold dark matter and hot dark matter,  probably exist in nature. This example illustrates that having two independent schools of thoughts was actually beneficial for progress  in cosmology. 
Paddy has taught us numerous times how  to think `outside the box'. We wish him many more years of original research in cosmology.

\begin{acknowledgement}
I thank my DES and UCL collaborators  for many inspiring discussions on this topic over the years, 
and for support from a European Research Council Advanced Grant FP7/291329.

\end{acknowledgement}
%

\input{Lahav_referenc}

\end{document}

%% file: Lahav_referenc.tex
%%%%%%%%%%%%%%%%%%%%%%%% referenc.tex %%%%%%%%%%%%%%%%%%%%%%%%%%%%%%
% sample references
% %
% Use this file as a template for your own input.
%
%%%%%%%%%%%%%%%%%%%%%%%% Springer-Verlag %%%%%%%%%%%%%%%%%%%%%%%%%%
%
% BibTeX users please use
% \bibliographystyle{}
% \bibliography{}
%